\documentclass[aps,prl,twocolumn,groupedaddress]{revtex4}
\usepackage{graphicx}
\usepackage{amssymb}
\usepackage{epstopdf}
\usepackage{color}

\begin{document}

\title{Rapid Single-Shot Measurement of a Singlet-Triplet Qubit
}

\author{C. Barthel$^1$}
\author{D. J. Reilly$^{1,2}$}
\author{C. M. Marcus$^1$}
\author{M. P. Hanson$^3$}
\author{A. C. Gossard$^3$}
\affiliation{$^1$Department of Physics, Harvard University, 17 Oxford Street, Cambridge, Massachusetts 02138, USA\\
$^2$School of Physics, University of Sydney, Sydney, 2006, Australia\\
$^3$Materials Department, University of California, Santa Barbara, California 93106, USA}


\begin{abstract}
We report repeated single-shot measurements of the two-electron spin state in a GaAs double quantum dot. The readout scheme allows measurement with fidelity above $90\%$ with a $\sim 7~\mu$s cycle time. Hyperfine-induced precession between singlet and triplet states of the two-electron system are  directly observed, as nuclear Overhauser fields are quasi-static on the time scale of the measurement cycle. Repeated measurements on millisecond to second time scales reveal evolution of the nuclear environment. \end{abstract}

\pacs{}

\maketitle
Qubits constructed from spin states of confined electrons are of interest for quantum information processing~\cite{lossPRA98}, for investigating decoherence and controlled entanglement, and as probes of mesoscopic nuclear spin environments. For logical qubits formed from pairs of electron spins in quantum dots~\cite{Levy02},  several requirements for quantum computing~\cite{divincenzocrit}  have been realized \cite{petta05,nowack2007,ladriere08,ReillyZamboni}. To date, however, measurements of these systems have constituted ensemble averages over time, while protocols for quantum control, including quantum error correction, typically require high-fidelity single-shot readout. Coherent evolution conditional on individual measurement outcomes can give rise to interesting non-classical states~\cite{armen2002,Romito2008}. Rapidly repeated single-shot measurements can also give access to the dynamics of the environment, allowing, for instance, feedback-controlled manipulation of the nuclear state. Single-shot measurements of solid-state quantum systems have been reported for superconducting qubits~\cite{Astafiev04}, the charge state of a single quantum dot~\cite{Lu2003}, the spin of a single electron in a quantum dot in large magnetic fields~\cite{elzerman04b,Amasha06}, and the two-electron spin state in a single quantum dot~\cite{MeunierPRB06}. 

In this Letter, we demonstrate rapidly repeated high-fidelity single-shot measurements of a two-electron spin (singlet-triplet) qubit in a double quantum dot. Singlet and triplet spin states are mapped to charge states~\cite{petta05}, which are measured by a radio-frequency quantum point contact~(rf-QPC) that is energized only during readout. The measurement integration time required for $> 90\%$ readout fidelity is a few microseconds. On that time scale, nuclear Overhauser fields are quasi-static, leading to observed periodic precession of the qubit. By measuring over longer times, the evolution of the Overhauser fields from milliseconds to several seconds can be seen as well.  We apply a model of single-shot readout statistics that accounts for $T_1$ relaxation, and find good agreement with experiment. Finally, we examine the evolution of the two-electron spin state at the resonance between the singlet ($S$) and the $m = +1$ triplet ($T_+$) via repeated single-shot measurement, and show that the transverse component of the Overhauser field difference is {\em not} quasi-static on the 
time scale of data acquisition, as expected theoretically.
\begin{figure}[b]
\includegraphics[width=3.0in]{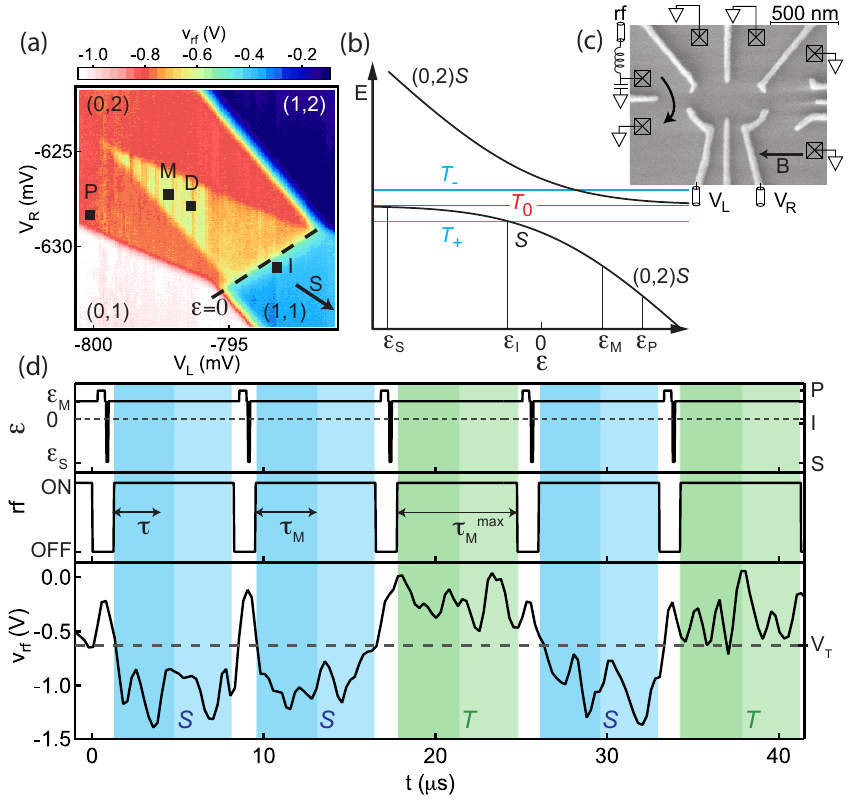}%
\caption{\label{Fig1}(a)~Charge occupancy (left, right) of the double dot, detected using rf-QPC reflectometer voltage, $v_{\rm rf}$, in continuous-sensing mode~\cite{ReillyZamboni} rather than single-shot readout.  The yellow triangle in (0,2) indicates where charge state (1,1) is metastable. Markers indicate gate voltages used in single-shot mode. Preparation of $(0,2)$ singlet (P); separation for $S - T_0$ mixing (S) and $S-T_+$ mixing (I); measurement (M); operating point with $0\,$V pulse amplitude (D). (b)~Two-electron energy levels as a function of detuning $\epsilon$ from (0,2) - (1,1) degeneracy.  
(c)~Micrograph of device identical to measured device, indicating ohmic contacts (boxes), fast gate lines, reflectometry circuit, grounded contacts, and field direction. (d)~Pulse-sequence of  $\epsilon$, controlled by $V_{\rm{R}}$ and $V_{\rm{L}}$, cycling through the points P, S, M. Sensor signal $v_{\rm{rf}}$ indicates triplet ($T$) or singlet ($S$) outcome for $\tau_{\rm{S}} =100$~ns. Integration subinterval time $\tau_{\rm{M}}$ chosen in post-processing.}
\end{figure}

The double quantum dot is formed by Ti/Au depletion gates on a GaAs/Al$_{0.3}$Ga$_{0.7}$As heterostructure with a two-dimensional electron gas  (density $2\times10^{15} ~\rm{m}^{-2}$, mobility $20~\rm{m}^2$/Vs) 100 nm below the surface. In order to split the three triplets, an in-plane magnetic field, $B$, larger than the typical Overhauser fields is applied along the line between dot centers. Except where noted, $B = 200$~mT.
As described elsewhere~\cite{Reillyapl07}, a proximal radio-frequency quantum point contact (rf-QPC) is sensitive to the charge state of the double dot, yielding an output signal $v_{{\rm{rf}}}$ via reflectometry, with sub-microsecond time resolution.    The charge state of the double dot is controlled by fast-pulsed gate voltages $V_{\rm{L}}$ and $V_{\rm{R}}$ from two synchronized Tektronix AWG710B arbitrary waveform generators. 

Energy levels of the system as a function of detuning, $\epsilon$, from the (1,1) - (0,2) charge degeneracy (controlled by $V_{\rm{R}}$ and $V_{\rm{L}}$) are shown in Fig.~1(b). The qubit comprises the two-electron singlet ($S$) and $m=0$ triplet ($T_0$) of the (1,1) charge state \cite{petta05}.  A pulse cycle [Fig.~1(d)] first prepares a spin singlet in $(0,2)$ by waiting at point P (near the edge of (0,2)) for $\tau_{\rm{P}} = 400$~ns, then moving to a separation point S~(I), where $S$ and $T_0$ ($S$ and $T_+$) are nearly degenerate, for a time $\tau_{\rm{S}} $~($\tau_{\rm{I}}$). Finally the system is brought to the measurement point M for a time $\tau_{\rm{M}}^{\rm{max}}$. 
If the separated electrons are in a singlet configuration when the system is pulsed to M, the system will return to $(0,2)$, which will be detected by the rf-QPC. If the two electrons are in a triplet state, they will remain in $(1,1)$ at point M, and detected accordingly. Coherent superpositions will be projected to the corresponding charge state during measurement. The rf-QPC is only energized during read-out, at point M [Fig.~1(d)].

The rf-QPC conductance is $\sim5\%$ higher in (0,2) than in (1,1), yielding a charge sensitivity of $6\times10^{-4}$~e/Hz$^{-1/2}$, i.e., unity signal-to-noise after $400$~ns of integration. To increase fidelity, single-shot outcomes are averaged over a sub-interval  $\tau_{\rm{M}}$ of the full measurement time $\tau_{\rm{M}}^{\rm{max}}$,  $V_{{\rm{rf}}} = 1/\tau_{{\rm{M}}}\int_0^{\tau_{\rm{M}}} v_{{\rm{rf}}}(\tau)d\tau $. By designating a threshold voltage $V_{\rm{T}}$, outcomes can be classified as singlet for $V_{{\rm{rf}}}<V_{\rm{T}}$ or triplet otherwise. Optimization of $\tau_{\rm{M}}$ and $V_{\rm{T}}$ is described below.

\begin{figure}
\includegraphics[width=3in]{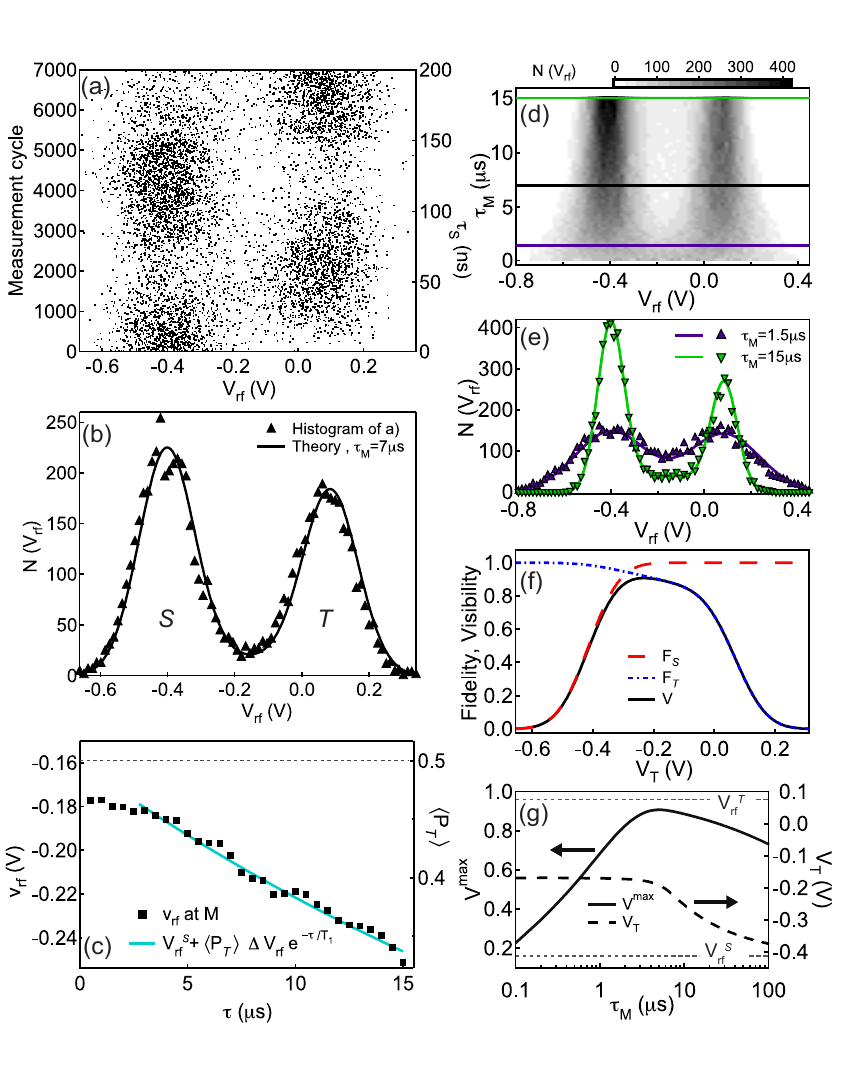}%
\caption{\label{Fig2} (a) 7000 consecutive single-shot measurements of $V_{{\rm{rf}}}$ using the pulse sequence in Fig.~1(d) with integration subinterval $\tau_{\rm{M}}=7~\mu$s and separation time, $\tau_{\rm{S}}$, incremented every 200 cycles~\cite{backgroundslope}. (b)~Histogram of the outcomes in (a), along with model (solid curve) \cite{freeparam}. (c) Instantaneous rf-QPC output $v_{{\rm{rf}}}(\tau)$ at time $\tau$ following pulsing to M, averaged over all cycles, along with a fit to the model, giving $T_1 = 34\,\mu$s \cite{freeparam}. (d)~Histograms $N(V_{\rm{rf}})$ (grayscale) for varying $\tau_{\rm{M}}$. (e)~Horizontal cuts through d) along with model, with values of the parameters {V$_{\rm{rf}}^{S}$}, V$_{\rm{rf}}^{T}$  from a fit to the $\tau_{\rm{M}}=15~\mu$s data~\cite{freeparam}.  (f)~Fidelity of singlet, $F_S $, and triplet, $F_T$, and visibility  $V  = F_S + F_T -1$ as a function of threshold, $V_{\rm{T}}$, for data in b).  (g)~Maximum visibility, $V^{\rm{max}}$, and optimal threshold, $V_{\rm{T}}$, as a function of measurement time, $\tau_{\rm{M}}$.
}
\end{figure}
Figure 2(a) shows 7000 consecutive one-shot measurements of the $S - T_0$ qubit with $\tau_{\rm{S}}$ ranging from 1 -- 200~ns, stepped by $\sim 6$~ns every 200 cycles. For these data, the integration subinterval,  $\tau_{\rm{M}}=7~\mu$s, was roughly half of the full measurement time, $\tau_{\rm{M}}^{\rm{max}} = 15~\mu$s.
The histogram of single-shot outcomes (Fig.~2(b)), with voltage bin width $V_{\rm{bin}} \sim10$~mV, is bimodal, with one peak at $V_{\rm{rf}}^{S}$, corresponding to the singlet ((0,2) charge state) outcome, and the other peak at $V_{\rm{rf}}^{T}$, corresponding to the triplet ((1,1) charge-state) outcome. The splitting $\Delta V_{\rm{rf}} = V_{\rm{rf}}^{T}-V_{\rm{rf}}^{S}$ reflects the difference in output of the rf-QPC between (0,2) and (1,1) charge states, while the width, $\sigma$, of the two peaks reflects measurement noise \cite{Reillyapl07}. However, the histogram is not simply the sum of two noise-broadened gaussians, because some states in (1,1) decay (with relaxation time $T_1$~\cite{Johnson05b}) during the measurement subinterval.  We model the full histogram $N(V_{\rm{rf}}) = N_{\rm{tot}}[n_{S}(V_{\rm{rf}})  + n_{T}(V_{\rm{rf}})] V_{\rm{bin}}$ as the sum of probability densities of singlet outcomes, $n_{S}(V_{\rm{rf}})$, and triplet outcomes, $n_{T}(V_{\rm{rf}})$, with $N_{\rm{tot}}$ the total number of measurements. The singlet probability density is modeled as a noise-broadened gaussian around $V_{\rm{rf}}^{S}$, 
\begin{equation}
\label{Histmodelcontrib}
n_{S}(V_{\rm{rf}})  =(1- \langle{P_{T}}\rangle)\,e^{-\frac{(V_{\rm{rf}}-V_{\rm{rf}}^{S})^2}{2\sigma^2}}\frac{1}{\sqrt{2\pi}\sigma},
\end{equation}
where $\langle{P_{T}}\rangle$ is the triplet probability over all $N_{\rm{tot}}$ outcomes. Triplet outcomes, on the other hand, can take on values spread between $V_{\rm{rf}}^{S}$ and  $V_{\rm{rf}}^{T}$ (and beyond, including measurement noise) to account for relaxation during the subinterval $\tau_{\rm{M}}$, 

\begin{eqnarray}
\label{TimeResolved}
\nonumber n_{T}(V_{\rm{rf}})  =e^{-\tau_{\rm{M}}/T_1}\langle{P_{T}}\rangle\, e^{-\frac{(V_{\rm{rf}}-V_{\rm{rf}}^{T})^2}{2\sigma^2}}\frac{1}{\sqrt{2\pi}\sigma} \\
+ \int_{V_{\rm{rf}}^{S}}^{V_{\rm{rf}}^{T}}\frac{\tau_{\rm{M}}}{T_1}\frac{\langle{P_{T}}\rangle}{\Delta V_{\rm{rf}}}e^{-\frac{V-V_{\rm{rf}}^{S}}{\Delta V_{\rm{rf}}}\frac{\tau_{\rm{M}}}{T_1}}e^{-\frac{(V_{\rm{rf}}-V)^2}{2\sigma^2}}\frac{dV}{\sqrt{2\pi}\sigma}.
\end{eqnarray}

The $T_1$ relaxation of the (1,1) triplet can be measured directly from the instantaneous rf-QPC output, $v_{\rm{rf}}(\tau)$, as a function of time $\tau$ following pulsing to point M (Fig.~2(c)). A fit of the ensemble-averaged rf-QPC output to the exponential form  $v_{\rm{rf}}(\tau) = V_{\rm{rf}}^{S} + \langle{P_{T}}\rangle\Delta V_{\rm{rf}}\, e^{-\tau / T_1}$
yields $\langle{P_{T}}\rangle = 0.5$ 
and $T_1 = 34\, \mu s$, using values for $V_{\rm{rf}}^{S}$ and $\Delta V_{\rm{rf}}$ determined from a fit  of the $N(V_{\rm{rf}})$ model to the $\tau_{\rm{M}} = 15~\mu$s histogram data \cite{width}. 

The tradeoff for optimizing the integration subinterval $\tau_{\rm{M}}$ is evident in Fig.~2(d), which shows histograms for a range of $\tau_{\rm{M}}$ from 0.25~$\mu$s to 15~$\mu$s. For short $\tau_{\rm{M}}$, the two peaks are blurred due to measurement noise; for long $\tau_{\rm{M}}$, the triplet peak loses strength due to relaxation.  To optimize the readout, we first define fidelities, $F_S$ and $F_T$, of pure singlet ($P_T=0$) and pure triplet state ($P_T=1$),
\begin{eqnarray}
\label{Fidelities}
F_S = 1 - \int_{V_{\rm{T}}}^{\infty} n_{S}(V) dV, \,\, F_T = 1 - \int_{-\infty}^{V_{\rm{T}}} n_{T}(V) dV,
\end{eqnarray}
following Ref.~\cite{elzerman04b}.
The integrals in Eq.~(\ref{Fidelities}) are the probabilities of misidentifying a pure singlet as a triplet and vice versa.
Figure~2(f) shows these fidelities as well as the visibility, $V= F_S + F_T -1$, for the $\tau_{\rm{M}} = 7~\mu$s data [from Fig.~2(b)] as a function of the threshold voltage $V_{\rm{T}}$. For this value of  $\tau_{\rm{M}}$, the maximum visibility, $\sim 90\%$, is achieved for $V_{\rm{T}}$ slightly less than the mean of $V_{\rm{rf}}^{T} $ and $V_{\rm{rf}}^{S} $ so that a triplet decaying towards the end of $\tau_{\rm{M}}$ still gets counted correctly. 
Optimal thresholds, $V_{\rm{T}}$, along with their associated maximum visibilities, $V^{\rm{max}}$, are plotted in Fig.~2(g) as a function of $\tau_{\rm{M}}$ using experimentally determined values for $T_1$, $V_{\rm{rf}}^{T} $, $V_{\rm{rf}}^{S}$, and $\sigma(\tau_{\rm{M}})$ \cite{freeparam, width}. The highest visibility, $\gtrsim90\%$, is realized for $\tau_{\rm{M}} \sim 6~\mu$s.

\begin{figure}
\includegraphics{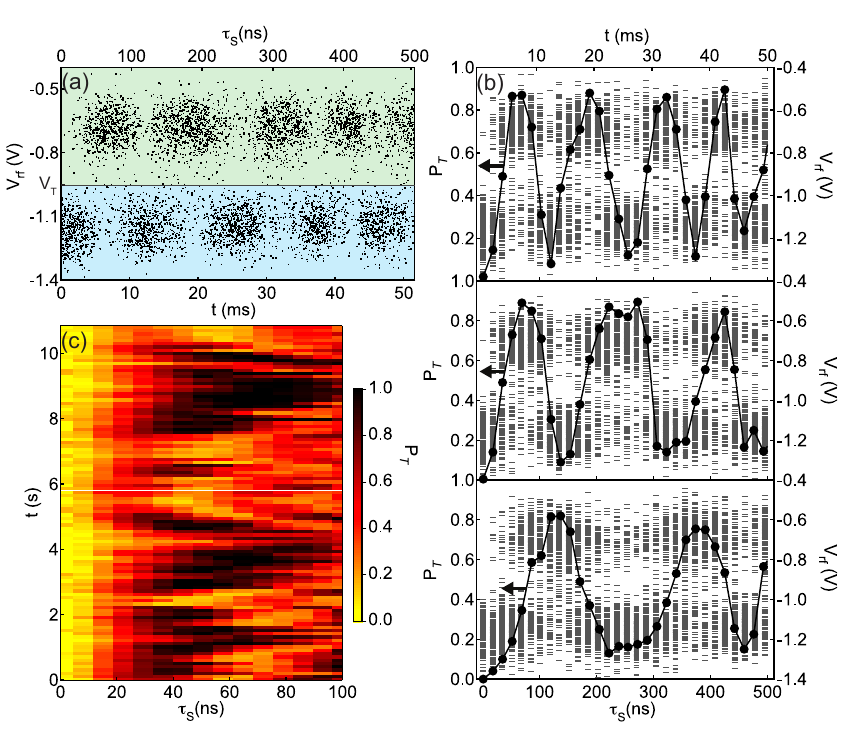}%
\caption{\label{Fig3}(a)~6000 consecutive single-shot $S - T_0$ measurements, $V_{\rm{rf}}$, with separation times $\tau_{\rm{S}}$ stepped by $\sim 17$~ns every 200 cycles, as a function of overall measurement time, $t$ (bottom axis). Threshold $V_{\rm{T}}$ separates outcomes identified as singlet (blue) or triplet (green). Oscillations due to Overhauser fields are evident, with slightly evolving period.  (b)~Single-shot outcomes (gray markers) and triplet probabilities, $P_{T}$, (black circles) over $\tau_{\rm{S}}$, for three nominally identical runs taken 10 minutes apart. 
(c)~Rapid acquisition of 108 $P_{T}$ traces at times $t$.  Probabilities $P_{T}$ are determined from  400 measurements per $\tau_{\rm{S}}$. 
}
\end{figure}
Previous work using continous charge sensing showed inhomogenous dephasing of the $S - T_0$ qubit, which was attributed to precession with a broad frequency spectrum, driven by the fluctuating Overhauser field difference between the two dots~\cite{petta05,TaylorPRB2007}.  For sufficiently fast single-shot repetition, Overhauser fields remain quasi-static over many single-shot measurements, leading to periodic $S - T_0$ precession, as seen in Fig.~3(a). Also evident is a variation of the precession period over $\sim 50$~ms, reflecting the slowly evolving nuclear environment, consistent with previous measurement and theory \cite{ReillyCorr07}. 

Variation in the $S - T_0$ precession period is more clearly demonstrated in Figs.~3(b,c). Figure~3(b) shows three sets of precession data taken 10 minutes apart. Periods of the oscillating triplet probability, $P_T$, defined by the average of 400 binary outcomes (either $S$ or $T_0$), correspond to longitudinal Overhauser field differences $\Delta B_{\rm{z}}^{\rm{nuc}} =1.3, 1.1, 0.4$~mT (top to bottom).  The continuous evolution of the nuclear environment can be seen in Fig.~3(c), which shows $P_T$ as a function of separation time $\tau_{\rm{S}}$---each row comparable to a panel in Fig.~3(b), but for $\tau_{\rm{S}}$ up to 100 ns rather than 500 ns---in slices taken every 100 ms \cite{corrdetails}. The meandering light-dark pattern reflects the random evolution of the $S-T_0$ precession period on a $\sim1\,$s time scale, consistent with dipole-dipole mediated nuclear diffusion~\cite{ReillyCorr07}. 

Assembling $P_T(\tau_{\rm{S}})$ statistics from single-shot measurements as a function of separation time $\tau_{\rm{S}}$ yields a time-averaged curve from which an inhomogeneous dephasing time $T_2^*$ can be extracted \cite{petta05,TaylorPRB2007}. Each point in Fig.~4(a) is an average over 1600 triplet-state return probabilities, each derived from 400 binary single-shot measurements. The individual $P_{T}(\tau_{\rm{S}})$ measurements are separated in time by $\sim20$ s. A fit to the theoretical gaussian form, $P_{T}(\tau_{\rm S}) = P_{T}(0) + (V/2)[1 - e^{-({\tau_{\rm{S}}/T_2^*})^2}]$, yields $T_2^* = 27$~ns, consistent with previous results \cite{petta05, ReillyCorr07}, visibility $V = 0.28$, and intercept $P_{T}(0) = 1 - F_S = 0.08$. These values yield a reasonable singlet fidelity, $F_S = 0.92$, but relatively low triplet fidelity $F_T = 0.36$ for this data (compared to Figs.~2,3) due to a short $T_1$ in this run.

\begin{figure}
\includegraphics{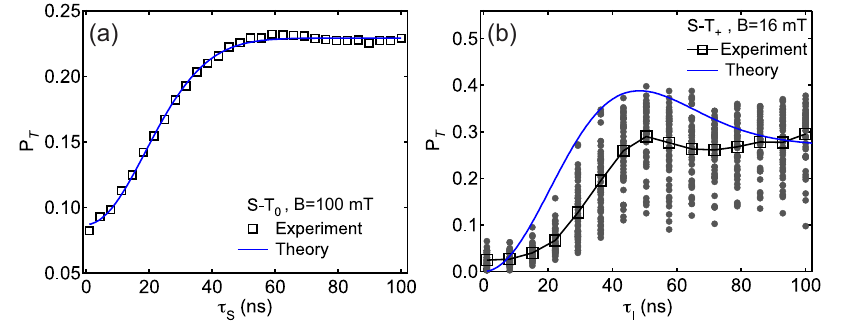}
\caption{\label{Fig4}(a)Triplet probability $P_{T}$  as a function of separation time $\tau_{\rm{S}}$ based on 1600 single-shot measurements per separation time. Fit (blue curve) gives $T_2^* = 27$~ns, $P_{T}(0) = 1-F_S = 0.08$, $V = 0.28$ (see text). (b) $P_{T}$ as a function of separation time $\tau_{\rm{I}}$ to the $S-T_+$ anticrossing at point I. Probabilities (gray circles) based on 400 single-shot binary measurements. At each $\tau_{\rm{I}}$, 50 values of $P_{T}$ are taken 15~s apart and averaged (open black squares). Theory curve, Eq.~(6), yields $B^{\rm nuc} = 0.7$~mT ($T_2^* = 37$~ns), with relatively poor agreement between experiment and theory (see text).}
\end{figure}

Finally, we investigate the triplet probability $P_{T}$ after separating (0,2) singlets to the point I, where the (1,1) singlet state $S$ crosses the $T_+$ triplet [see Fig.~1(b)]. Whereas mixing of $S$ and $T_0$ at point S relies on the component of the Overhauser field difference {\em along} the total field direction, mixing of $S$ and $T_+$ at point I relies on the component of the Overhauser field difference {\em transverse} to the total field. Evolution of transverse Overhauser fields are not inhibited by nuclear or electron Zeeman energy differences, and is relatively fast, set by nuclear dipole-dipole ($\sim 100~\mu$s) and Knight-shift ($\sim 10~\mu$s) energetics ~\cite{TaylorPRB2007, Jakethesis, ReillyCorr07, Hanson2007}. As expected, we do not observe periodic precession between $S$ and $T_+$.  We note a variation over the course of the measurement in spin-flip probability at a fixed $\tau_{\rm{I}}$ and separation point I. This is likely due to changes in the position of the narrow $S-T_+$ resonance resulting from a small build-up of nuclear polarization during the measurement \cite{Jakethesis, Petta2008}.

Figure~4(b) shows probabilities $P_{T}$ for the $T_+$ state as a function of $\tau_{\rm{I}}$.  Each probability value (gray circle) in Fig.~4(b) is based on 400 binary single-shot measurements with $\tau_{\rm{M}}=8~\mu\rm{s}$. Series of $P_{T}(\tau_{\rm{I}})$ measurements were made over a range of $\tau_{\rm{I}}$ up to $ \sim 100\,$ns, with an acquisition time $\sim 50$~ms per series. A total of 50 series, spaced by $\sim 15$~s to allow decorrelation of longitudinal Overhauser fields, were then averaged to give the black squares in Fig.~4(b).  
 
 At small external fields, when the $S-T_+$ anticrossing is in (1,1), the probability of detecting a triplet following separation for a time $\tau_{\rm{I}}$  can be written $P_T = P_T^0 + V \int d^3\mathbf{B}\, \rho(\mathbf{B}) [{(\Delta B_{\rm{x}}^2+ \Delta B_{\rm{y}}^2)/2 (\hbar \omega / |g^* \mu_B|)^2}]\sin^2(\omega \tau_{\rm{I}})$, where $\omega = |g^*\mu_B|/(2\hbar)[B_{\rm{z}}^2+2(\Delta B_{\rm{x}}^2+\Delta B_{\rm{y}}^2)]^{1/2}$ is the precession rate between $S$ and $T_+$ at the center of the anticrossing, $\Delta B_{\rm{x(y)}} =[B^{L}_{\rm{x(y)}}-B^{R}_{\rm{x(y)}}]/2$ are transverse Overhauser field differences between left (L) and right (R) dots, $B_{\rm{z}} =[B^{L}_{\rm{z}}+B^{R}_{\rm{z}}]/2$ is the average longitudinal Overhauser field, $V$ is readout visibility, and $g^*=-0.44$ is the effective electron g factor in GaAs.  Assuming Overhauser fields $\mathbf{B}=(\Delta B_{\rm{x}}, \Delta B_{\rm{y}}, B_{\rm{z}})$ are gaussian distributed on long time scales,  $\rho(\mathbf{B}) = (2\pi B^{\rm{nuc}})^{-3/2}e^{-(\mathbf{B}/ B^{\rm{nuc}})^2 /2}$, yields the form in Fig.~4(b) \cite{integralform}.  Setting $P_T^0=1-F_S=0$ and $B^{\rm nuc}$ to match the overshoot in the data yields $B^{\rm nuc}=0.7$~mT $\sim \hbar|g^* \mu_B T_2^*|^{-1}$, corresponding to $T_2^* \sim 40$~ns, and $V = F_T \sim 0.7$. Unlike Fig.~4(a), theory and experiment do not match well for the $S-T_+$ mixing, due in part to the the buildup of average nuclear polarization, which shifts the $S-T_+$ resonance and lowers $P_T$. 

\begin{acknowledgments}
We acknowledge support from IARPA/ARO, and the Department of Defense. Devices were made at Harvard's Center for Nanoscale Systems (CNS), a member of the NSF National Nanotechnology Infrastructure Network (NNIN). We thank J. M. Taylor and E. A. Laird for useful discussion, and J. Waissman for technical assistance.
\end{acknowledgments}

\bibliographystyle{apsrev}

\begin{thebibliography}{20}
\bibitem{corrdetails} Each $P_T$ value (pixel) in Fig.~3(c) is based on 1000 single-shot measurements. Each row of 15 probabilities is acquired in 80~ms, followed by 20~ms of dead time. For these measurements, $B = 100$ mT.
\end{thebibliography}

\end{document}